\documentclass[aps,prb,11pt,groupedaddress]{revtex4}  
\usepackage{graphicx}  
\usepackage{subfigure}  
\usepackage{setspace}  
\usepackage{dcolumn}   
\usepackage{bm}        
\usepackage{amssymb}   
\usepackage{natbib}
\begin{document}
\singlespacing
\title{Quantitative analysis of chain packing in polymer melts using large scale molecular dynamics simulations}

\author{{\bf Rajeev Kumar\footnote[1]{To whom any correspondence should be addressed, Email : kumarr@ornl.gov}}}

\affiliation{National Center for Computational Sciences, Oak Ridge National Laboratory, Oak Ridge, TN 37831}
\author{{\bf Bobby G. Sumpter}}

\affiliation{Center for Nanophase Materials Science, Oak Ridge National Laboratory, Oak Ridge, TN 37831}

\begin{abstract}

\noindent 
\textbf{Abstract:} We have carried out a quantitative analysis of the chain packing in polymeric melts using molecular 
dynamics simulations. The analysis involves constructing Voronoi tessellations 
in the equilibrated configurations of the polymeric melts. In this work, we focus on the effects of 
temperature and polymer backbone rigidity on the packing. We found that the Voronoi polyhedra near the chain 
ends are of higher volumes than those constructed around the other sites along the backbone. 
Furthermore, we demonstrated that the backbone rigidity (tuned by fixing the bond angles) affect 
the Voronoi cell distribution in a significant manner, especially at lower temperatures. 
For the melts consisting of chains with fixed bond angles, the Voronoi cell distribution was found to 
be wider than that for the freely jointed chains without any angular restrictions. 
As the temperature is increased, the effect of backbone rigidity on the Voronoi cell distributions 
diminishes and becomes similar to that of the freely jointed chains. Demonstrated dependencies of the distribution 
of the Voronoi cell volumes on the nature of the polymers are argued to be important for efficiently 
designing the polymeric materials for various energy applications. 
\end{abstract}

\maketitle

\section{Introduction}
Polymers have been proposed\cite{polymer_electrolyte_book,conjugated_poly,hoppe_review,capacitors} as an 
alternative materials for providing mechanical flexibility and stability at relatively low cost for myriad energy applications. 
Such application include lithium batteries\cite{polymer_electrolyte_book}, 
organic photovoltaics\cite{conjugated_poly,hoppe_review}, and supercapacitors\cite{capacitors}. 
However, there is considerable scope when it comes to designing and synthesizing efficient polymer 
materials for energy applications despite a plethora of research both by the experimental and theoretical community.

A fundamental understanding of how polymer chains pack in the melts is of paramount importance for the development 
of a molecular description of the transport\cite{polymer_electrolyte_book,sokolov_macro} 
of electrolytes through the polymer matrix. Structure-property concepts such as fragility\cite{sokolov_macro} demands a systematic 
study of the chain packing effects and the free-volume distribution\cite{rigby_roe,douglas_voronoi}
in polymeric melts. In general, the chain packing depends on 
the architecture of the polymers and several other experimental variables affecting crystallinity of the materials.
However, in the amorphous state, two key experimental variables are the temperature and the backbone rigidity, which 
clearly affect the performance of 
different polymeric materials in devices\cite{polymer_electrolyte_book,conjugated_poly,hoppe_review}. 

In this work, we have studied the effects of temperature and backbone rigidity on the packing 
of chains in amorphous melts using molecular dynamics (MD) simulations. The simulation method is presented in
 section ~\ref{sec:method} followed by the results and conclusions in sections ~\ref{sec:results} and ~\ref{sec:conclusions}, 
respectively. 
\section{Simulation method}\label{sec:method}
We have used the Large-scale Atomic/Molecular Massively Parallel Simulator\cite{lammps} (LAMMPS) 
for carrying out the Brownian dynamics (BD) simulations for the polymer melts. The polymer chains are modeled by a 
coarse-grained model (known as Kremer-Grest bead-spring model\cite{kremer_grest}).  
In particular, the Langevin equation is integrated, given by 
\begin{eqnarray}
m\ddot{\mathbf{r}}_i &=& -\zeta \dot{\mathbf{r}}_i - \nabla_{\mathbf{r}_i}U\left(\left\{\mathbf{r}_i\right\}\right) + f_i(t) ,
\end{eqnarray}
where $m,\zeta$ are the mass and the friction coefficient for the beads. $U\left(\left\{\mathbf{r}_i\right\}\right)$ is 
the pairwise interaction potential and is given by 
$U\left(\left\{\mathbf{r}_i\right\}\right) = U_{\mbox{LJ}}\left(\mathbf{r}_{ij}\right) + U_{\mbox{FENE}}\left(\mathbf{r}_{ij}\right)$, 
where $U_{\mbox{LJ}}$ is the truncated and shifted Lennard-Jones potential, which is purely \textit{repulsive}. Explicitly, it 
is given by

\[ U_{\mbox{LJ}}\left(\mathbf{r}_{ij}\right)  = \left\{ \begin{array}{ll}
         4\epsilon\left[\left(\frac{\sigma}{r_{ij}}\right)^{12} - \left(\frac{\sigma}{r_{ij}}\right)^{6}  + \frac{1}{4}\right], 
& \mbox{$r_{ij} \leq 2^{1/6} \sigma$}\\
        0, & \mbox{$r_{ij}  \geq 2^{1/6} \sigma$}.\end{array} \right. \] 

Furthermore, the finite extensible nonlinear elastic (FENE) potential is used in combination with 
the Lennard-Jones potential to 
maintain the topology of the molecules. Explicitly, $U_{\mbox{FENE}}$ can be written as 
\[ U_{\mbox{FENE}}\left(\mathbf{r}_{ij}\right)  = \left\{ \begin{array}{ll}
         -0.5 k r_c^2 \ln \left[1 - \left(\frac{r_{ij}}{r_c}\right)^{2} \right], 
& \mbox{$r_{ij} \leq r_c$}\\
        \infty, & \mbox{$r_{ij}  \geq r_c$}.\end{array} \right. \]
so that $r_c$ is the maximum extent of a bond. Here, we have used the notation $r_{ij} = |\mathbf{r}_{ij}|$. 
Following Ref. \cite{kremer_grest}, we have chosen $r_c = 1.5 \sigma, k = 30 \epsilon/\sigma^2, \zeta = 0.1/\tau$, 
where $\tau = \sigma \sqrt{m/\epsilon}$ 
is the parameter used to quantify time steps. Results presented in this paper were obtained by carrying out the simulations 
in dimensionless units so that $r_c^\star = r_c/\sigma = 1.5 , k^\star = k\sigma^2/\epsilon = 30, \zeta^\star = \zeta \tau = 0.1$ and 
$\epsilon^\star = 2 \epsilon$ (used in $U_{\mbox{LJ}}$ replacing $\epsilon$) so that explicit 
values of $\epsilon, \sigma$ and $m$ are needed to switch from dimensionless units to 
the real units. 

In this work, we present results for polymer melts corresponding to a number density of monomers 
$\rho^\star = \rho \sigma^3 = 0.85$, which 
corresponds to $128,000$ particles in a cubic box of edge length $L^\star = L/\sigma = 53.20$. 
The simulations are performed by using periodic boundary conditions for two sets of chains: freely jointed chains and the chains with fixed 
angles between adjacent bonds. For the chains with fixed bond angles, we have used the \textit{harmonic} angular potential 
to constrain the angle between adjacent bonds. Explicitly, the angular potential used in this work is 
written as 
\begin{eqnarray}
U_{H} &=& \frac{K_\theta}{2}\left[\theta - \theta_o\right]^2.
\end{eqnarray}
Simulation results presented here were obtained by choosing $K_\theta = 20 \epsilon$ and $\theta_0 = 2\pi/3$ (in radians), which corresponds to a 
bond angle of $120$ degrees. Polymer chains with $64$ beads are studied at five 
temperatures $T^\star = k_B T/\epsilon = 0.2, 0.4, 0.6, 0.8$ and $1$. 
Furthermore, all the simulations are carried out in the NVT ensemble so that the number density 
stays the same during the simulation runs. In other words, the effects of thermal expansion and the change in 
the number density with the change in temperature are not considered here.
All the simulations were started from random 
configurations of the chains followed by a slow push-off set-up run for one million time steps as described in Ref. ~\cite{kremer_grest} before 
moving on to the runs with the interaction potentials described above.  

To quantify the chain packing, we have carried out a Voronoi analysis\cite{voronoi_book,voronoi_code} for the polymeric melts 
after fifty million time steps (in units of $\tau$). 
The Voronoi construction is carried out using the C++ library, Voro++. The results presented here 
are the averages of one thousand configurations spanning 1 million time steps, and we have carefully ensured that the 
results are independent of subsequent runs. Furthermore, the MD simualtions have been carried out by using $2,400$ 
cores on the supercomputer, Jaguar, followed by the Voronoi analysis on the data analysis cluster, Lens. 
 
\section{Results}\label{sec:results}
\begin{figure}[ht]  \centering
\subfigure[]{\includegraphics[height=2.25in]{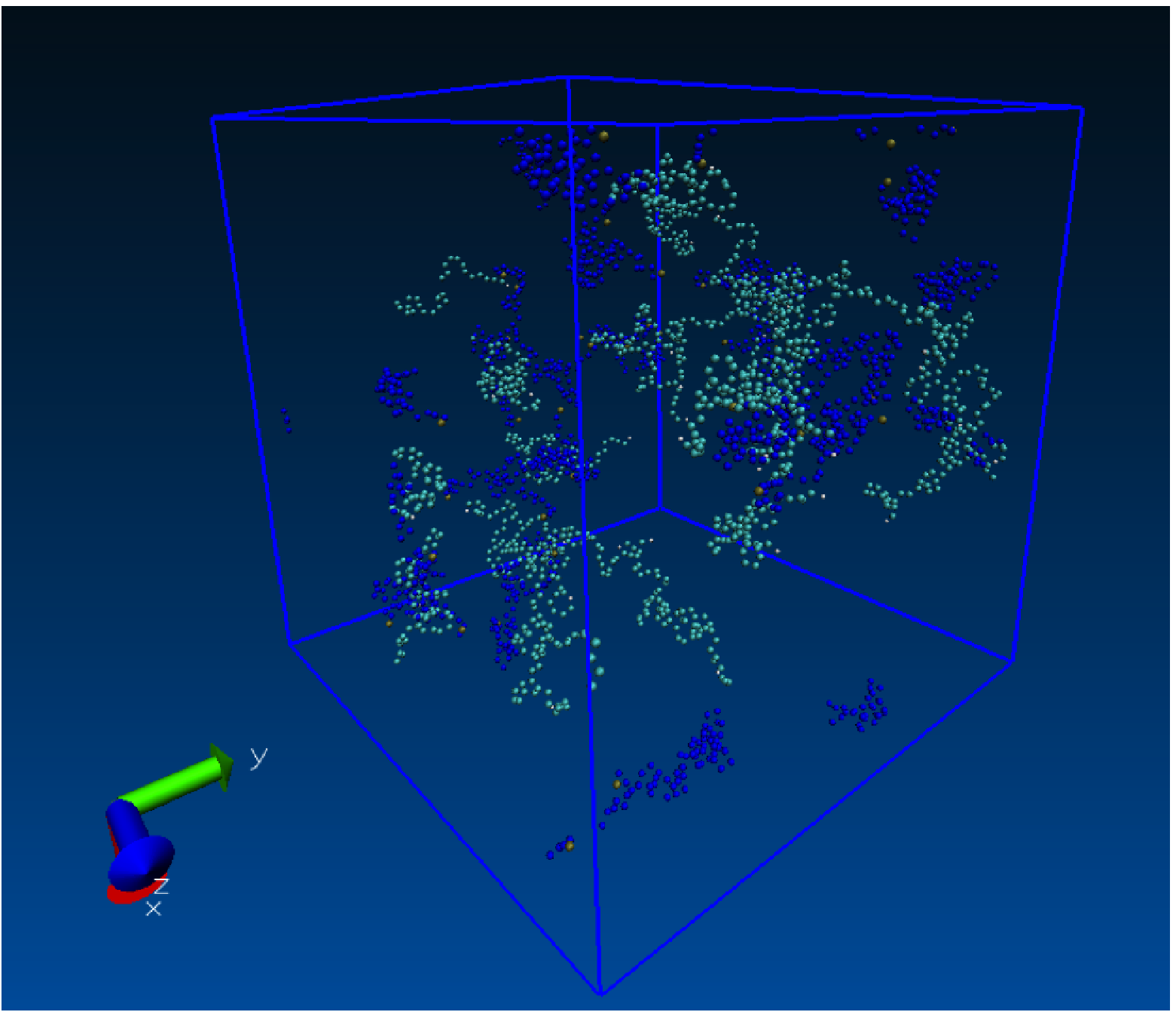}}
\subfigure[]{\includegraphics[height=2.25in]{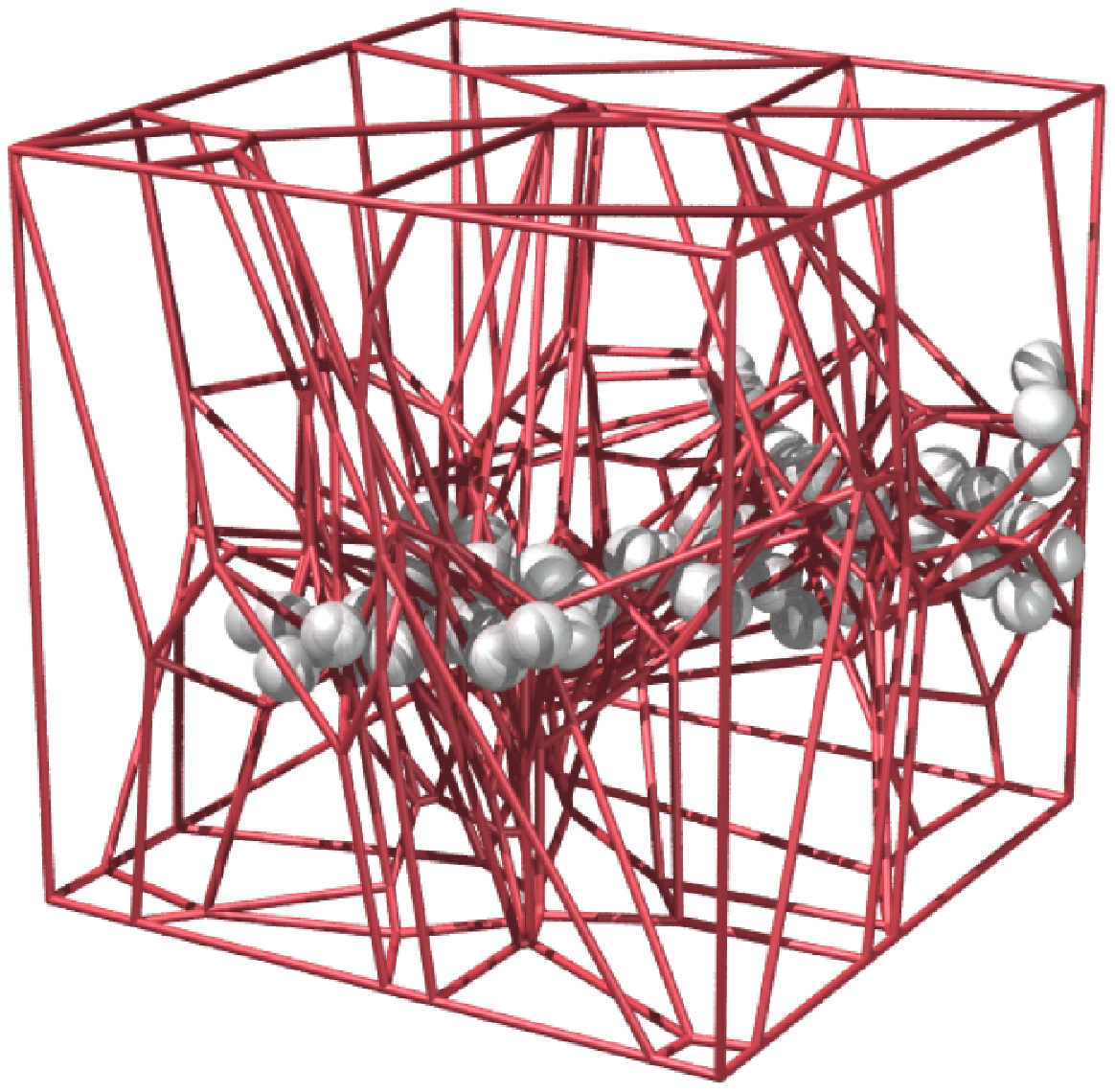}}
\caption{ Conformations of a couple of chains in the melts (containing $2,000$ chains) are shown in Fig. ~\ref{fig:snapshots}(a). The chains are colored differently for 
demonstration purposes. In Fig. ~\ref{fig:snapshots}(b) a Voronoi construction for a single polymer chain in a cubic box is shown.}
\label{fig:snapshots}
\end{figure}

A snapshot of the polymer chains in the simulation box is shown in Fig. ~\ref{fig:snapshots} along with the Voronoi construction for 
a single chain in a cubic box. It is clear from Fig. ~\ref{fig:snapshots}(b) that the shape and volume of the Voronoi polyhedra are highly 
sensitive to the conformations of the chain. In the case of the melts, the conformational degrees of freedom of the chains depends on the 
intra- and interchain interactions in a complicated manner. The BD simulation procedure described above allows us to 
study the effect of these interactions on the distribution of the Voronoi polyhedra in the melts and quantify the chain packing. 

\begin{figure}[ht]
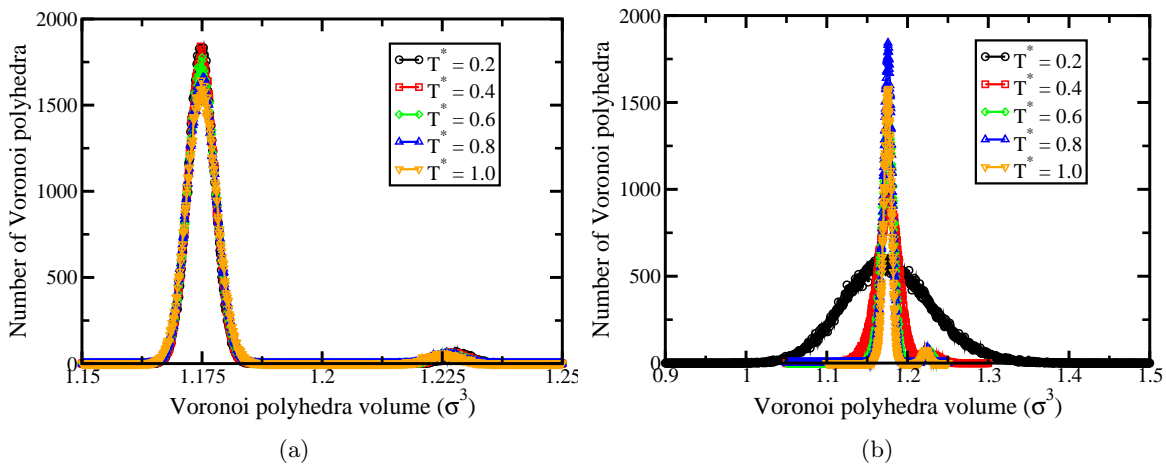
  \centering
\subfigure[]{\includegraphics[width=3.0in]{N64_f_allT.eps}} 
\subfigure[]{\includegraphics[width=3.0in]{N64_sf_allT.eps}}
\caption{Distribution of the Voronoi polyhedra volume in the melts containing (a) freely jointed chains 
and (b)  the chains with fixed bond angles. The distribution is similar for the two cases for higher temperatures and 
significantly different at lower temperatures (e.g., $T^\star = 0.2$).}
\label{fig:voro_compare}
\end{figure}

In Fig. ~\ref{fig:voro_compare}, we present the histograms for the distribution of the Voronoi polyhedra in the 
melts containing the freely jointed chains and the chains with the fixed bond angles at different temperatures. 
These results, obtained for a fixed volume of the simulation box, clearly show that the temperature affects the 
distribution of the Voronoi polyhedra in a significant manner. At higher temperatures corresponding 
to $T^\star = 0.6, 0.8, 1.0$, 
there are two sets of populations for both the freely jointed chains and the chains with the fixed bond angles. 
For a number density of $\rho^\star = 0.85$, the average volume per particle is $1/0.85 =  1.176$ (in units of $\sigma^3$). 
Indeed a clear peak at the polyhedra volume of $1.176$ is seen at all the temperatures with a finite width. An additional peak at 
higher volumes is observed and is found to correspond to the chain ends. Higher Voronoi polyhedra volume associated with the 
chain ends has been reported in the literature\cite{rigby_roe}.  

However, at lower temperatures (e.g., at $T^\star = 0.2$) there are significant differences in the distibution of the 
Voronoi polyhedra, when the case of the freely jointed chains is compared with the chains with fixed bond angle. In the case of the 
chains with fixed bond angles, which correspond to relatively rigid chains, the distribution is wider in comparison with the 
freely jointed chains.\\ 
 
\section{Conclusions} \label{sec:conclusions}
In summary, we have demonstrated that polymer backbone rigidity affects the distribution of the free-volume 
in the polymeric melts, especially at low temperatures. In contrast to the fully flexible chains such as 
in PEO, rigid chains have a wider distribution of the free-volume at lower temperatures. With the 
assumption that the distibution of the free-volume affects\cite{polymer_electrolyte_book} the transport of electrolytes, these results are important 
for an efficient design of polymeric materials intended for use in energy applications. In the case of organic photovoltaics, 
conjugated polymers (i.e., polymers with rigid backbones) are already used\cite{conjugated_poly,hoppe_review}. However, flexible macromolecules such as PEO are 
used for battery applications and show lower\cite{polymer_electrolyte_book} ionic conductivity for room-temperature applications. A wider distribution of 
the free-volume at lower temperatures 
is indeed necessary for the materials to conduct. Keeping this in mind, our results show that the polymeric materials with 
rigid backbone needs to be used for low-temperature applications, in contrast to the widely used 
approach of PEO-based materials. Also, we point out that this relatively newer idea of using rigid chains for battery applications 
conforms to some recently published experimental results\cite{sokolov_macro}. 

\section*{Acknowledgments} \label{acknowledgement}
This research used resources of the Oak Ridge Leadership Computing
Facility at the Oak Ridge National Laboratory, which is supported by the
Office of Science of the U.S. Department of Energy under Contract No.
DE-AC05-00OR22725. Also, we acknowledge the financial support from UT-Battelle 
through LDRD program (project $\# 05608$).

\end{document}